\documentclass[runningheads]{llncs}
\usepackage{booktabs}
\usepackage{multirow}
\usepackage{array} 
\usepackage{algorithmic}
\usepackage{algorithm}
\usepackage[T1]{fontenc}
\usepackage{marvosym}
\usepackage{ifsym}
\usepackage{graphicx}
\begin{document}
\title{High Capacity Reversible Data Hiding for Encrypted 3D Mesh Models Based on Topology\thanks{Supported by National Natural Science Foundation of China (62172001, 61872003).}}
\titlerunning{High Capacity Reversible Data Hiding for Encrypted 3D Mesh Models}

\author{Yun Tang\inst{1}\and Lulu Cheng\inst{1}\and Wanli Lyv\inst{1}\and Zhaoxia Yin\inst{2}\textsuperscript{(\Letter)}\orcidID{0000-0003-0387-4806} }

\authorrunning{Y. Tang et al.}
%
\institute{ Anhui Province Key Laboratory of Multimodal Cognitive Computation, School of Computer Science and Technology, Anhui University, 230601, PR China  \and
School of Communication \& Electronic Engineering, East China Normal University, Shanghai 200241, China}
%
\maketitle              
\begin{abstract}
Reversible data hiding in encrypted domain(RDH-ED) can not only protect the privacy of 3D mesh models and embed additional data, but also recover original models and extract additional data losslessly. However, due to the insufficient use of model topology, the existing methods have not achieved satisfactory results in terms of embedding capacity. To further improve the capacity, a RDH-ED method is proposed based on the topology of the 3D mesh models, which divides the vertices into two parts: embedding set and prediction set. And after integer mapping, the embedding ability of the embedding set is calculated by the prediction set. It is then passed to the data hider for embedding additional data. Finally, the additional data and the original models can be extracted and recovered respectively by the receiver with the correct keys. Experiments declare that compared with the existing methods, this method can obtain the highest embedding capacity.
\end{abstract}
\section{Introduction}
Data hiding\cite{barton1997method} is a method to achieve copyright protection and covert communication by embedding additional data into the cover media and extracting the data without errors. However, the original media can not be restored losslessly by the traditional data hiding methods. It is a fatal problem in the military, medical and other fields where data integrity is strictly required. To solve this problem, reversible data hiding(RDH) is proposed.

RDH can embed additional data in the cover media and restore both media and data without errors. Over the years, researchers have devised many methods. They can be divided into three main categories: lossless compression\cite{fridrich2002lossless}, difference expansion\cite{tian2003reversible} and histogram shifting\cite{ni2006reversible}. In lossless compression methods, some features of the original media are discovered and losslessly compressed, making room for storing additional data. Difference expansion explores redundancy between adjacent pixels and embeds additional data into their difference. Histogram shifting embeds additional data by shifting peak points toward zero points in the pixel distribution histogram.

With the boom in cloud computing and privacy protection, many multimedia are uploaded, such as digital images, audio, 3D mesh models, etc., to the cloud for storage and transmission. In order to protect the privacy and security of data, media will be encrypted before uploading. In this scenario, RDH-ED has caught the attention of many researchers. There are two main categories of RDH-ED: vacating room after encryption (VRAE)\cite{yin2014separable,zhang2011reversible,zhang2011separable} and reserving room before encryption (RRBE)\cite{ma2013reversible,puteaux2018efficient,yin2019reversible}. Zhang et al.\cite{zhang2011reversible} introduced a RDH scheme for encrypted images based on VRAE. The encrypted images were divided into non-overlapping blocks, and each block can store 1 bit of data by flipping the 3 least significant bits(LSB) of the encrypted data. In \cite{zhang2011separable}, a separable RDH-ED method was proposed. In this method, a specific matrix multiplication was designed to losslessly compress the LSB of the encrypted image, which implemented the separability of data extraction and image restoration. However, the embedding ability of these methods is low due to encrypted image entropy maximization. To increase the embedding capacity and reduce the bit error rate, Ma et al.\cite{ma2013reversible} first proposed a method based on RRBE, which implemented both data extraction and image recovery losslessly based on histogram shifting. Although many RDH-ED methods have obtained considerable outcomes, there is not enough research on 3D mesh models.

As an emerging digital media following images, audio and videos, 3D mesh models are applied in many fields, such as medical treatment, construction, video games, etc. To protect the privacy security and verification copyright of 3D mesh models in massive application scenarios, it is essential to study the RDH-ED methods of 3D mesh models. Jiang et al.\cite{jiang2017reversible} first proposed the RDH method based on VRAE for 3D mesh models in encrypted domain. However, the data extracted may be errors and the capacity of this method is not satisfactory. Shah et al.\cite{shah2018homomorphic} and van Rensburg et al.\cite{van2021homomorphic} applied homomorphic encryption to RDH-ED method and embedded data in dual domain. Although homomorphic encryption increases the capacity of the method, the computational cost is high. Tsai\cite{tsai2020separable} proposed a native method by spatial subdivision and space encoding. However, If the parameters are not chosen properly, errors may occur during extraction. Xu et al.\cite{xu2022separable} proposed the most significant bit(MSB) prediction, which implemented extraction and recovery separably. In \cite{yin2021separable}, the embeddable room is extended to multi-MSB. However, the vertices are not fully utilized. Lyu et al.\cite{lyu2022high} made the use of vertices more fully by adding labels. However, the topology of 3D mesh models is not  utilized and the division of vertices is not sufficient. Therefore, this study proposes a vertex division method based on the model topology and manages to the highest embedding capacity compared with state-of-the-art methods. 

The main contributions of the proposed method are as follows:

1)We take full advantage of the topology of the 3D mesh models and propose a more reasonable vertex division method.

2)Our division method enables the correlation between vertices to be further exploited. Experiments declare that the proposed method has the highest embedding capacity compared to other methods.

The remainder of this paper is organized as follows: The related works are presented in Section \ref{relatedworks}. Section \ref{proposedmethod} presents the proposed method. The experimental results and analysis are shown in Section \ref{results}. Finally, the paper is concluded in Section \ref{concusion}.
\section{Related Works}\label{relatedworks}
In this section, we introduce the types of research models and the existing RDH-ED methods for 3D mesh models.

In computer-aided design, 3D mesh models are usually represented as triangular or polygonal models, which are composed of vertex data and face data. This paper studies the most popular triangular 3D mesh models. Vertex data includes vertex coordinates and is represented as $V=\{v_i|1 \leq i \leq n\}$ where $n$ donates the number of vertices. In the rectangular coordinate system of space, each vertex has three values in the $x$, $y$, and $z$ directions. Topological relationships between vertices are contained in face data, which is represented as $F=\{f_1,f_2,...,f_m\}$ and $m$ is the number of the face. Each face $f$ contains three vertices. The corresponding file format is shown in Table \ref{format}.
\begin{table}[]
	\centering
	\caption{File format of a triangular mesh model.}\label{format}
	\begin{tabular}{cccccc}
		\toprule
		\multicolumn{4}{c}{Vertex data}                           & \multicolumn{2}{c}{Face data}                \\ 
		Index of vertex &\ $x$-axis       &\ $y$-axis        &\ $z$-axis          &\ Index of face  & \ Elements of each face              \\ \hline
		$v_{1}$        & $v_{1,x}$    & $v_{1,y}$   & $v_{1,z}$   & $f_{1}$       & ($v_{15}, v_{2}, v_{1}$) \\ 
		$v_{2}$        & $v_{2,x}$    & $v_{2,y}$   & $v_{2,z}$   & $f_{2}$        & ($v_{4}, v_{6}, v_{1}$) \\ 
		...             & ...         & ...         & ...         & ...            & ...                       \\ 
		$v_{31}$        & $v_{31,x}$  & $v_{31,y}$  & $v_{31,z}$  & $f_{31}$    & ($v_{6}, v_{1}, v_{3}$) \\ 
		...             & ...         & ...         & ...         & ...           & ...                         \\ 
		\bottomrule
	\end{tabular}
	\centering
\end{table}

Jiang et al.\cite{jiang2017reversible} introduced the RDH method to 3D mesh models in encrypted domain for the first time. The vertices were first mapped to integers and encrypted by stream cipher. After encryption, additional data was embedded into the multi-LSB. However, this method is not satisfactory in terms of embedding capacity and there may be errors in data extraction. Shah et al.\cite{shah2018homomorphic} proposed a two-tier RDH-ED framework. The authentication information was embedded by the data sender in the first tier and additional data was embedded by the cloud server in the second tier. This method improves the embedding capacity by applying homomorphic encryption, but this method brings the file size increments and the encryption process is expensive. Van Rensburg et al.\cite{van2021homomorphic} applied Paillier encryption to 3D mesh models and proposed an improved two-layer RDH method. This method minimizes file size increments as much as possible and improves the capacity. However, the original mesh models cannot be restored losslessly. Tsai\cite{tsai2020separable} proposed a separable RDH method based on spatial subdivision and space encoding for encrypted 3D mesh models. However, this method has the problem of bit error during extracting data. Xu et al.\cite{xu2022separable} used the MSB prediction and proposed a new method based on RRBE, which increased the embedding capacity. Vertices were divided into embedding set and reference set by traversing the face data. The reference set is used for predicting the embedding set. After division and MSB prediction, the additional data was embedded in the MSB plane of the embedding set. Based on \cite{xu2022separable}, Yin et al.\cite{yin2021separable} implemented higher embedding capacity by utilizing multi-MSB to hide data. However, this method does not adaptively assign payload to the vertices of the embedding set. Lyu et al.\cite{lyu2022high} introduced vertex labels as auxiliary information and improved the utilization of vertices. However, this method has shortcomings in the division of vertices. This method only divided the vertices by the index of vertex data and ignored the face data. To divide the vertices more reasonably, we propose a vertex division method based on model topology, which has much improved the embedding capacity compared with state-of-the-art methods.
\begin{figure}
	\includegraphics[width=\textwidth]{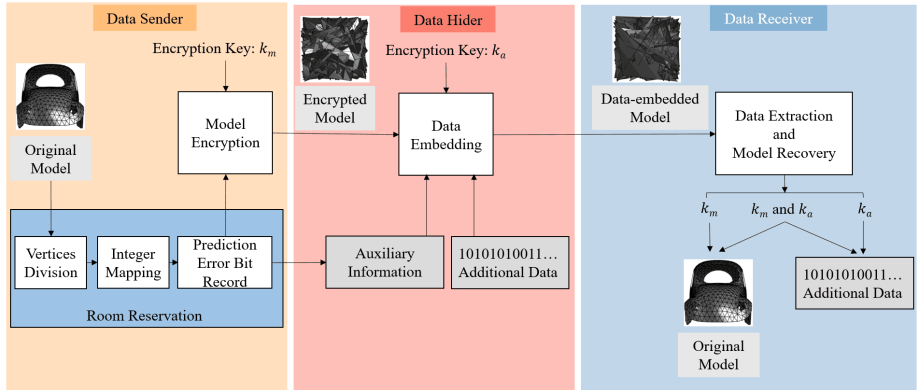}
	\caption{Method flowchart.} \label{flowchart}
\end{figure}
\section{Proposed Method}\label{proposedmethod}
In this section, the proposed method is described in detail. As shown in Fig.\ref{flowchart}, the proposed method consists of four parts: (1)The mesh models are preprocessed to make room for embedding additional data. (2)The 3D mesh models are encrypted by stream ciphers. (3)The encrypted additional data is embedded in the encrypted model by the data hider. (4)The additional data and original models can be extracted and recovered losslessly by the data receiver who has the corresponding key.   

\subsection{Room Reservation}\label{roomreservation}
\subsubsection{Vertices Division}
We propose a new division method and increase the number of the embedding vertices reasonably. In \cite{jiang2017reversible,xu2022separable,yin2019reversible}, the vertices were divided according to the face data. Each face was traversed in numbering order, and at most 1 vertex in each face was added to the embedding set. This method is not sufficient for the utilization of vertex data, and can only reach 33\% of the vertex utilization for embedding.  In \cite{lyu2022high}, the vertices were divided according to the parity of the vertex index. This method applied more vertices to embedding additional data and the vertex utilization ratio reached 50\%. However, only the vertex data was used to divide vertices. To divide the vertices more reasonably, a native division method combining vertex data and face data is proposed. 
\begin{figure}
	\includegraphics[width=\textwidth]{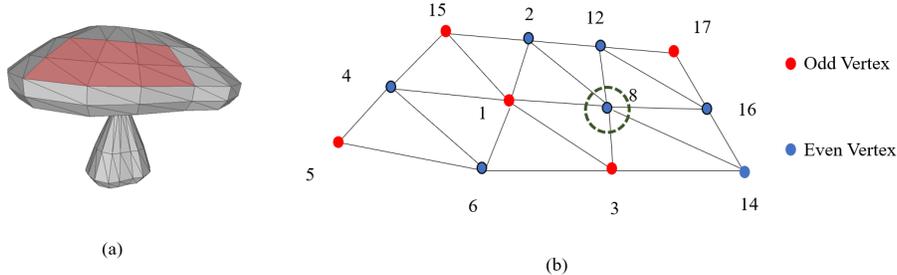}
	\caption{(a) Mushroom, (b) The labeled local topology of Mushroom.}\label{topology}
\end{figure}

In our method, the vertices are divided into two parts first according to the parity of the vertex index. For the convenience of introduction, here we select the even-numbered part as the prediction set $S_p$ and the odd-numbered part as the embedding set $S_e$. Some vertices in $S_p$ are redundant and unsuitable to use as prediction vertices.  We traverse the vertices in $S_p$ one by one and filter out the vertices that are suitable for embedding additional data. In order to ensure that some vertices in $S_p$ with excellent predictive ability are not classified into the $S_e$, vertices are added to the $S_e$ only when the number of odd-numbered vertices around the vertices does not exceed twice the number of even-numbered vertices. Besides, in order to ensure the prediction accuracy, the vertices must also be guaranteed that they can be predicted by at least 2 vertices. An example of the vertices screening process is shown in Fig.\ref{topology}. The vertex $v_8$ has four even neighbor vertices which means that this vertex can be accurately predicted with multi-MSB and the number of odd-neighbored vertices does not exceed twice the number of even-numbered vertices. Therefore, $v_8$ is more suitable for the $S_e$. After that, the predicted vertices for $v_1$ will have $v_8$ removed.

\subsubsection{Integer Mapping}
Uncompressed vertex coordinate values are stored in the computer as 32-bit floating point numbers. \cite{deering1995geometry}  suggested that 3D mesh models do not require this level of accuracy for most applications. Therefore, lossy compression is performed on the values of vertices. The value of vertices is expanded to between $-10^p$ and $10^p$, where $p \epsilon (1,33)$. The expanded value is processed into an integer by
\begin{equation}
	v_{i,j}' = \lfloor v_{i,j}\times 10^{p} \rfloor,\ i=1,2,...,n,\ j\epsilon (x,y,z),
\end{equation}
where $\lfloor . \rfloor$ is the round down function. The receiver can restore the coordinates to floating point coordinates by
\begin{equation}\label{recover double}
	v_{i,j}'' = v_{i,j}'/10^{p}, \  i=1,2,...,n,\ j\epsilon (x,y,z).
\end{equation}
The bit length $l$ of the integer is determined by the value of $p$, and the conversion relationship between them can be expressed as
\begin{equation}
	l = \left\{
	\begin{array}{l}
		8,\ 1\leq p\leq 2\\
		16,\ 3\leq p\leq 4\\
		32,\ 5\leq p\leq 8\\
		64,\ 10\leq p\leq 33.
	\end{array}
	\right. 
\end{equation}
\begin{figure}
	\includegraphics[width=\textwidth]{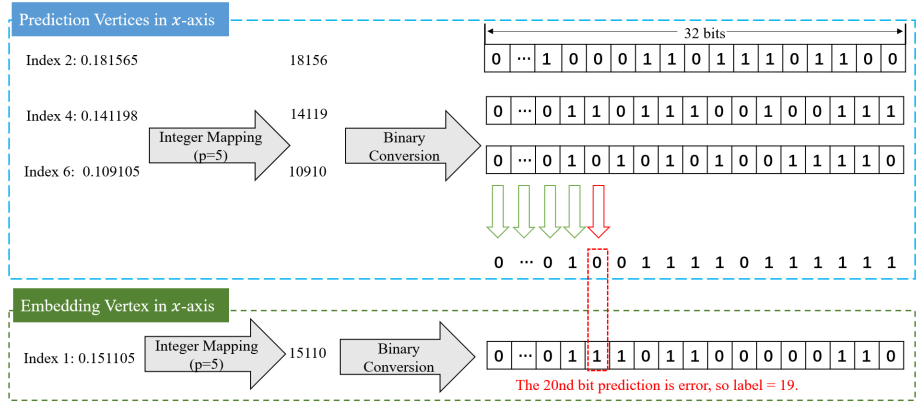}
	\caption{Prediction error labels for $v_1$ in $x$-axis. }\label{prediction}
\end{figure}
\subsubsection{Prediction Error Bit Record}
The embedding vertex is predicted by the vertices around it. In Fig.\ref{prediction}, the process of predicting error labels is illustrated in detail. The $v_1$ has three neighbor vertices in $S_p$, which are $v_2$,$v_4$ and $v_6$. From the MSB to the LSB, the predicted value of each bit plane is determined according to the mode of all corresponding bit planes of the prediction vertices. If the number of 0 and 1 is the same, 1 is chosen as the final value. The embeddable length is obtained by comparing the predicted value with $v_1$ and records as a label. Since a vertex has coordinates in three directions, the label selection for a vertex will finally be the minimum of the three directions. The embeddable length of each vertex is recorded in the location map, which is compressed by arithmetic coding and then embedded into the 3D mesh models as auxiliary information. 

\subsection{Model Encryption}
The 3D mesh model is encrypted by stream cipher in this paper. First, the coordinate of each vertex is converted to a binary representation. Each binary bit can be obtained by 
\begin{equation}\label{decimal2binary}
	b_{i,j,k} = \lfloor v_{i,j}'/2^{k}\rfloor\ {\rm mod}\ 2,\ k = 0,1,...,l-1.
\end{equation}
Then, the data sender uses the stream cipher function to generate pseudo-random bits $c_{i,j,k}$, and the encrypted bits are obtained by
\begin{equation}
	e_{i,j,k} = b_{i,j,k} \oplus c_{i,j,k}, \ k = 0,1,...,l-1,
\end{equation}
where  $\oplus$ means exclusive OR. Finally, the encrypted vertices can get by
\begin{equation}
	v_{i,j,k}''' = \sum^{l-1}_{k=0} e_{i,j,k} \times 10^k.
\end{equation}
\subsection{Data Embedding}
\begin{figure}
	\includegraphics[width=\textwidth]{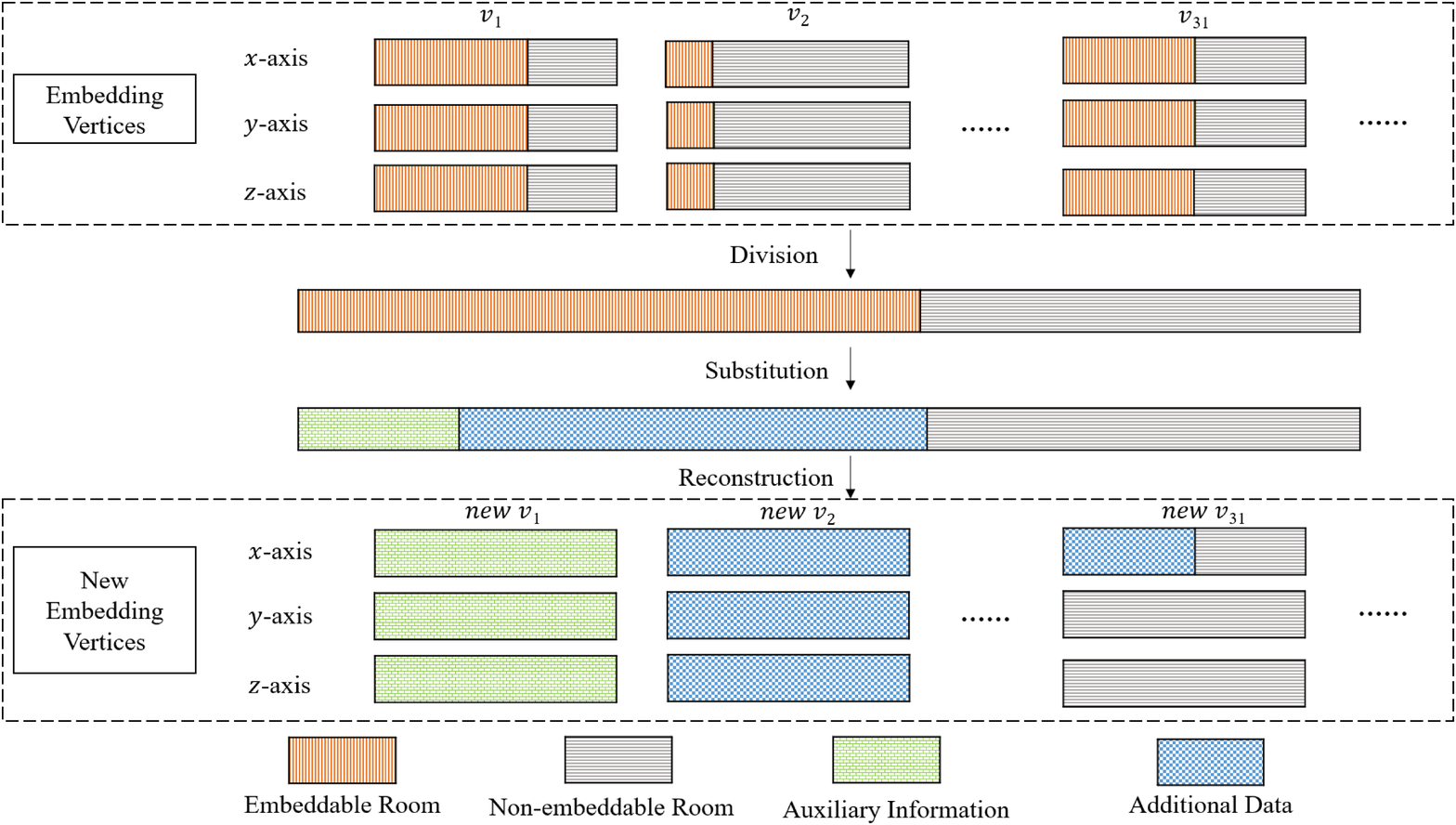}
	\caption{The process of data embedding.}\label{dataembedding}
\end{figure}

The process of data embedding is shown in Fig.\ref{dataembedding}. The room obtained in section\ref{roomreservation} is used for data hiding. First, the data hider divides the binary representation of vertex values in $S_{e}$ into two parts according to the location map, the embeddable part and the non-embeddable part. Then, the encrypted  additional data and auxiliary information are embedded into the embeddable part by substitution. Finally, the vertices are reconstructed.
\subsection{Data Extraction and Model Recovery}
The data receiver who has the correct key can decrypt the encrypted data. The additional data can be extracted with the key $k_a$ and original mesh models can be recovered with the key $k_m$. Therefore, here are three possible scenarios:

Case 1: The receiver only holds $k_a$. All the additional data can be extracted from the encrypted model and plaintext data is decrypted with $k_a$.

Case 2: The receiver only holds $k_m$. The data receiver extracts the auxiliary data and decrypts the mesh models through the key $k_m$. However, the embedding set in the decrypted mesh model is not the same as the original one. Then, the prediction set is applied to restore the embedding set with the help of the auxiliary data.

Case 3: The receiver holds both keys $k_a$ and $k_m$. At this time, the data receiver can obtain additional data and restore the original models at the same time. The procedure is the same as in case 1 and case 2, but model decryption is performed after data extraction.

\section{Experiment Results and Analysis}\label{results}
\subsection{Visual Quality and Quantitative Analysis}
\begin{figure}[ht]
	\includegraphics[width=\textwidth]{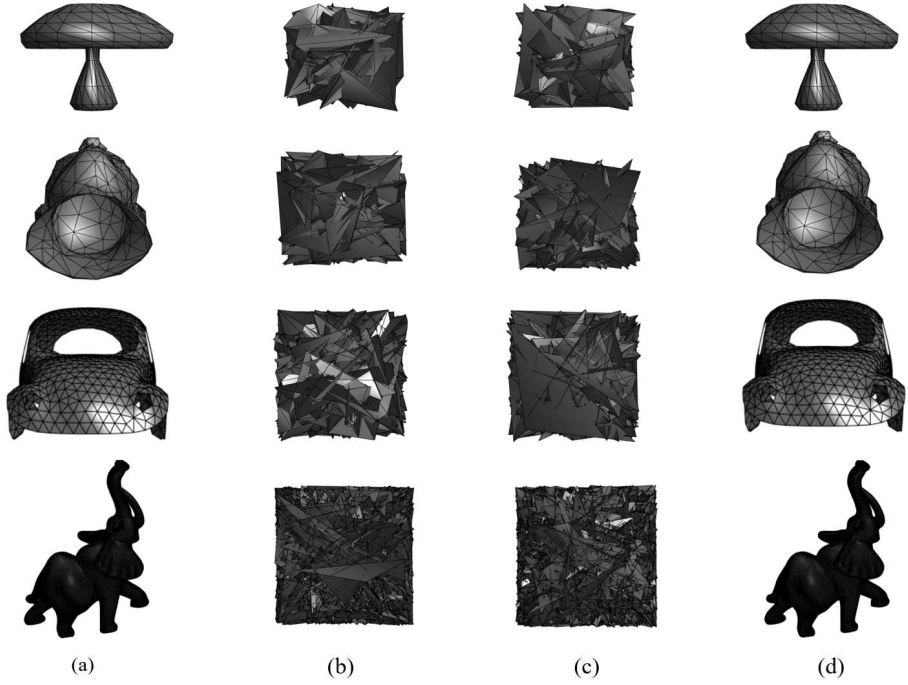}
	\caption{Models presentation at different stages($p$=5): (a) original models; (b) encrypted models; (c) data-embedded models; (d) recovery models.} \label{four_test_models}
\end{figure}
Four test mesh models are demonstrated in Fig.\ref{four_test_models}. Despite the slight distortion introduced to the models during the preprocess stage, there is little difference between the original models and the recovered models based on human visual observations. This shows that our method is feasible in the application. To further demonstrate the feasibility of the method, we introduce signal-to-noise ratio(SNR) and Hausdorff distance to quantitatively measure the gap between the original models and the recovered ones.

SNR can calculate the gap between two models. The larger the SNR, the smaller the gap between the two models. SNR can be obtained by 
\begin{equation}\label{eqSNR}
	{\rm SNR} = 10{\rm \log_{10}}\frac{\sum^N_{i=1}[(v_{i,x}-\bar{v}_{x})^{2}+(v_{i,y}-\bar{v}_{y})^{2}+(v_{i,z}-\bar{v}_{z})^{2}]}{\sum^N_{i=1}[(v_{i,x}''-v_{i,x})^{2}+(v_{i,y}''-v_{i,y})^{2}+(v_{i,z}''-v_{i,z})^{2}]},
\end{equation}
where $v_{i,x}, v_{i,y}, v_{i,z}$ are the values of original coordinates, $\bar{v}_{x}, \bar{v}_{y}, \bar{v}_{z}$ are the average values of original coordinates, and $v_{i,x}'', v_{i,y}'', v_{i,z}''$ are the values of recovery coordinates.

Hausdorff distance measures the maximum distance value among the closest distances between two model vertices. The smaller the Hausdorff distance, the smaller the gap between the two models. Assuming there are two sets $A=(a_1,a_2,...,a_u)$ and $B=(b_1,b_2,...,b_u)$, Hausdorff distance can be calculated as
\begin{equation}
		\begin{array}{l}
		D(A,B) = {\rm max}(d(A,B),d(B,A)),\\
		d(A,B) = \mathop{{\rm max}}\limits_{a\epsilon A}{\mathop{{\rm min}}\limits_{b\epsilon B}\left \|a-b\right \|},\\
		d(B,A) = \mathop{{\rm max}}\limits_{b\epsilon B}{\mathop{{\rm min}}\limits_{a\epsilon A}\left \|b-a\right \|},
		\end{array}
\end{equation}
where $\left \|.\right \|$ is the Euclidean distance, and $u$ denotes the number of elements in each set.
\begin{table}
	\centering
	\setlength{\tabcolsep}{1.5mm}{}
	\caption{SNR and Hausdorff distance of four test models($p$=5).}\label{SR}
	\begin{tabular}{ccccc}
		\toprule
		Test models & Number of vertex &Number of face & SNR     &Hausdorff distance($10^{-6}$) \\ \hline
		Mushroom    & 226           & 448         & 102.26  & 8.12     \\ 
		Mannequin   & 428           & 839         & 130.93  & 4.06     \\ 
		Beetle      & 988           & 1763        & 96.20   & 8.66     \\ 
		Elephant    & 24955         & 49918       & 95.97   & 8.66     \\ 
		\bottomrule
	\end{tabular}
	\centering
\end{table}

\cite{lyu2022high,xu2022separable,yin2021separable} have demonstrated that as the value of $p$ increases, although the distortion added into the models becomes smaller, the embedding capacity is lower. Under the trade-off of embedding capacity and the requirements of practical applications, the value of $p$ is to be 5. The results of four test models are shown in Table \ref{SR}. As can be seen in the table, the SNR of the four test models has reached a high value, and the Hausdorff distance is almost zero.
\subsection{Capacity Analysis}
The embedding rate(ER) is one of the important indicators of the RDH-ED methods for 3D mesh models in encrypted domain, which is presented by bit per vertex(bpv). ER can be calculated by
\begin{equation}
	{\rm ER} = \frac{l_p-l_{ai}}{n},
\end{equation}
where $l_p$ is the total capacity and $l_{ai}$ is the length of auxiliary information.
\begin{table}[]
	\centering
	\setlength{\tabcolsep}{2.5mm}{}
	\caption{The comparison of ER(bpv) on four test models.}\label{Capacity}
	\begin{tabular}{ccccccccc}
		\toprule
		Test models & \cite{jiang2017reversible} & \cite{shah2018homomorphic} & \cite{van2021homomorphic} & \cite{tsai2020separable} & \cite{xu2022separable}& \cite{yin2021separable} & \cite{lyu2022high} & \textbf{Ours} \\ \hline
		Mushroom  & 0.45    & 6       & 13      &7.68     &1.34     &16.72    &21.76   &\textbf{22.53}   \\ 
		Mannequin & 0.34    & 6       &13       &7.68     &0.95     &13.66    &18.05    &\textbf{24.08}\\ 
		Beetle    & 0.35    & 6       & 13      & 7.68    &0.98     &16.51    &23.55    &\textbf{31.75}\\ 
		Elephants & 0.34    &6        &13       &7.68     &1.02     &18.12    &27.96    &\textbf{38.93}   \\ 
		\bottomrule
	\end{tabular}
	\centering
\end{table}

The comparison of ER on four test models is shown in Table \ref{Capacity}. Jiang et al.\cite{jiang2017reversible} used multi-LSB for data embedding. The embedding rate of this method is not high enough, and it does not exceed 1 bpv at most. Shah et al.\cite{shah2018homomorphic} designed a two-layer embedding method. Additional data was embedded in the first layer through histogram expansion and shifting techniques. The second layer utilized the self-blinding property of the Paillier cryptosystem to embed additional data. The final embedding rate of this method is 6 bpv. Compared with \cite{shah2018homomorphic}, \cite{van2021homomorphic} adjusted the encryption bit length of the coordinates to reduce the file size increment. Tsai\cite{tsai2020separable} exploited the ratio of vertex value normalization for model encryption and data embedding and realized 7.68 bpv. Xu et al.\cite{xu2022separable} proposed to embed data in MSB. Although the embedding rate is improved, the relationship between the adjacent vertices has not been used fully. Yin et al.\cite{yin2021separable} extended the embedding range to multi-MSB and achieved a substantial increase in capacity. Lyu et al.\cite{lyu2022high} further improved the embedding capacity and improved the utilization of vertices by labeling vertices. Due to the topology of 3D mesh models, the same vertex can be included in multiple faces and associated with multiple vertices. Furthermore, many correct prediction bits can be obtained with the help of two or three vertices by rationally utilizing the strong multi-MSB correlation between adjacent vertices. However, among the aforementioned methods, the topological structure of the models is not used  to divide the vertices and the division of vertices is not reasonable enough. The proposed method addresses this issue and implemented the state-of-the-art results on the average ER. The division results of the embedded vertices and predicted vertices on four test models are displayed in Table \ref{divisionresults}. Except for the Mushroom model, the number of embedded vertices does not increase greatly after vertex division due to the small number of vertices and faces, the other three models all obtained a high vertex utilization ratio. The utilization of model topology increases the number of embedded vertices, and finally the ER of the models is improved.
\begin{table}
	\centering
	\setlength{\tabcolsep}{1mm}{}
	\caption{The results of vertices division on four test models($p$=5).}\label{divisionresults}
	\begin{tabular}{ccccc}
		\toprule
		\multirow{2}{*}{\begin{tabular}[c]{@{}c@{}}Test \\ models\end{tabular}} &
		\multirow{2}{*}{\begin{tabular}[c]{@{}c@{}}Number of \\ vertex\end{tabular}} &
		\multirow{2}{*}{\begin{tabular}[c]{@{}c@{}}Number of \\ face\end{tabular}} &
		\multirow{2}{*}{\begin{tabular}[c]{@{}c@{}}Number of \\ embedding vertex\end{tabular}} &
		\multirow{2}{*}{\begin{tabular}[c]{@{}c@{}}Vertex utilization \\ ratio(\%)\end{tabular}} \\
		&       &       &     &  \\ \hline
		Mushroom  & 226   & 448   & 121 & 53   \\
		Mannequin & 428   & 839   & 316    &73  \\
		Beetle    & 988   & 1763  & 705    &71  \\
		Elephant  & 24955 & 49918 & 18935    &75  \\
		\bottomrule
	\end{tabular}
	\centering
\end{table}

\begin{figure}[ht]
	\includegraphics[width=\textwidth]{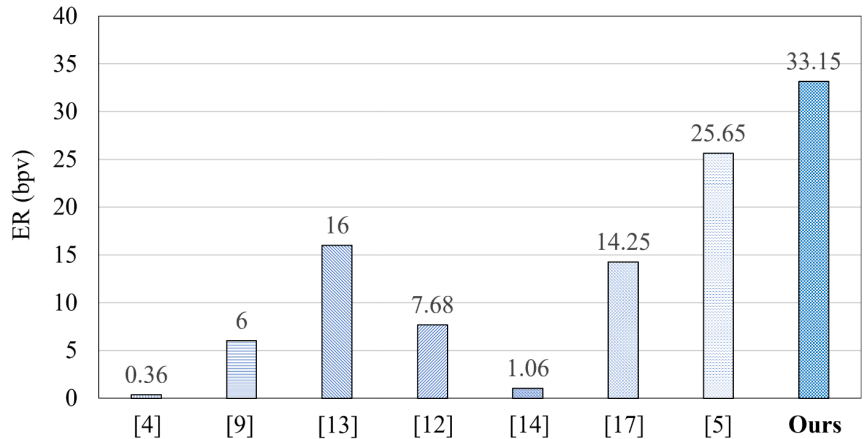}
	\caption{Average capacity comparison with other methods on the Princeton dataset.} \label{datasetcompare}
\end{figure}

To be more persuasive, the proposed method is tested on the Princeton dataset\cite{shilane2004princeton} which consists of 380 different 3D mesh models. The caparison with other methods is shown in Fig.\ref{datasetcompare} and our method also obtains state-of-the-art experimental results on the average ER.
\begin{table}[]
	\centering
	\caption{The comparison of features.}\label{features}
	\begin{tabular}{ccccc}
		\toprule
		Features & Encryption method        & Computation complexity & Separability & Data error \\ \hline
		\cite{jiang2017reversible}  & Stream cipher encryption & Low                    & No           & Yes        \\ 
		\cite{shah2018homomorphic}  & Homomorphic encryption   & High                   & No           & No         \\ 
		\cite{van2021homomorphic}  & Homomorphic encryption   & High                   & No           & No         \\ 
		\cite{tsai2020separable}  & Stream cipher encryption & Low                    & Yes          & Yes         \\ 
		\cite{xu2022separable}  & Stream cipher encryption & Low                    & Yes          & No         \\
		\cite{yin2021separable}  & Stream cipher encryption & Low                    & Yes          & No         \\ 
		\cite{lyu2022high}  & Stream cipher encryption & Low                    & Yes          & No         \\ 
		\textbf{Ours}     &\textbf{Stream cipher encryption} &\textbf{Low}                  &\textbf{Yes}       &\textbf{No}       \\ 
		\bottomrule
	\end{tabular}
	\centering
\end{table}
\subsection{Features Analysis}
The comparison of features with other methods is shown in Table \ref{features}. Jiang et al.\cite{jiang2017reversible} used the smoothness function, which would lead to errors in data extraction, making the model and data inseparable during recovery and extraction. In \cite{shah2018homomorphic,van2021homomorphic}, it is expensive to apply homomorphic encryption to the models and the computational complexity of these methods is very high. In \cite{tsai2020separable}, the selection of a suitable threshold is crucial, otherwise the data extraction will be wrong. \cite{lyu2022high,xu2022separable,yin2019reversible} are MSB-based methods that make room for embedding additional data by exploiting the correlations between vertices. The proposed method can not only recover the model and additional data reversibly, but also recover and extract them separably. Besides, the computational complexity of using stream cipher encryption is lower than that of homomorphic encryption. 
\section{Conclusion}\label{concusion}
In this paper, we study a RDH method for 3D mesh models  with high capacity in encrypted domain. The vertices are divided into two parts reasonably by utilizing the vertex data and face data. One is used for data embedding, the other one is used for model recovery. The experiments demonstrate that the proposed method reaches state-of-the-art performance. In the future, we consider to enhance the robustness of the method to resist channel interference of data during network transmission. Besides, a method that does not require the transmission of auxiliary information from the data sender to the data hider is also our next research direction.

\subsubsection{Acknowledgements} This research work is partly supported by National Natural Science Foundation of China (62172001, 61872003).
%
%
%
\bibliographystyle{splncs04}
\bibliography{ref}
\end{document}